\documentclass[aps,prl,reprint,groupedaddress]{revtex4-1}
\usepackage{amssymb}
\usepackage{amsmath}
\usepackage{array}
\usepackage{multirow,bigdelim}
\usepackage{enumitem}

\begin{document}

\title{Stochastic local operations and classical communication (SLOCC) and
local unitary operations (LU) classifications of n qubits via ranks and
singular values of the spin-flipping matrices}
\author{Dafa Li$^{1,2}$}

\begin{abstract}
We construct $\ell $-spin-flipping matrices from the coefficient matrices of
pure states of $n$ qubits and show that the $\ell $-spin-flipping matrices
are congruent and unitary congruent whenever two pure states of $n$ qubits
are SLOCC and LU equivalent, respectively. The congruence implies the
invariance of ranks of the $\ell $-spin-flipping matrices under SLOCC and
then permits a reduction of SLOCC classification of n qubits to calculation
of ranks of the $\ell $-spin-flipping matrices. The unitary congruence
implies the invariance of singular values of the $\ell $-spin-flipping
matrices under LU and then permits a reduction of LU classification of n
qubits to calculation of singular values of the $\ell $-spin-flipping
matrices. Furthermore, we show that the invariance of singular values of the
$\ell $-spin-flipping matrices $\Omega _{1}^{(n)}$ implies the invariance of
the concurrence for even $n$ qubits and the invariance of the n-tangle for
odd $n$ qubits. Thus, the concurrence and the n-tangle can be used for LU
classification and computing the concurrence and the n-tangle only performs
additions and multiplications of coefficients of states.
\end{abstract}


\affiliation{
$^1$Department of Mathematical Sciences, Tsinghua University,
Beijing, 100084, China\\
$^2$Center for Quantum Information Science and Technology, Tsinghua National
Laboratory for Information Science and Technology (TNList), Beijing,
100084, China\\}


\maketitle

\section{Introduction}

Entanglement is considered as a uniquely resource in quantum teleportation,
quantum cryptography and quantum information and computation \cite{Nielsen}.
There are different types of entanglement for multipartite systems.\ The
classification of entanglement plays an important rule in quantum
information theory \cite{Bennett}-\cite{Kraus2}. The following three types
of classifications of entanglement have widely been studied: LU \cite{Grassl}%
-\cite{Kraus2}, LOCC (local operations and classical communication) \cite%
{Gabor}, and SLOCC \cite{Bennett}, \cite{Dur}-\cite{LDFPRA15}.\

Considerable efforts have contributed to the SLOCC entanglement
classification. For example, the complete SLOCC classifications of two and
three qubits have been obtained. There are two (resp. six) SLOCC equivalence
classes for two (resp. three) qubits. For four qubits, there are infinite
SLOCC\ equivalence classes \cite{Dur} and the infinite SLOCC\ classes are
partitioned into nine inequivalent families\ \cite{Verstraete, Miyake,
Borsten, Viehmann}.\ It is known that a SLOCC classification for $n$ qubits
remains unsolved because the difficulty increases rapidly as $n$ does \cite%
{Bastin, Ribeiro,LDFPRL12,LDFPRA12,Sharma12,Gour, LDFPRA15}.

Recently, SLOCC invariant polynomials have been proposed for classification
of entanglement \cite{Luque}. Very recently, it has been shown that ranks of
the coefficient matrices of states are invariant under SLOCC and the
invariance of the ranks can be applied to SLOCC\ classification of
entanglement \cite{LDFPRL12, LDFPRA12, LDFPRA15, Shuhao}.

It is known that two LU equivalent states have the same amount of
entanglement and are equally used for any kind of application \cite{Kraus1,
Kraus2}. The polynomial invariants under LU were studied \cite{Grassl, Rains}%
. The necessary and sufficient conditions for LU equivalence of states of n
qubits were presented \cite{Kraus2} and\ used for LU classifications of two
to five qubits \cite{Kraus1}.

In this paper, we construct the $\ell $-spin-flipping matrices\ from the
coefficient matrices of pure states of $n$ qubits and prove that the $\ell $%
-spin-flipping\ matrices are congruent and unitary congruent under SLOCC and
LU, respectively. The congruence confirms the invariance of ranks and the
unitary congruence asserts the invariance of singular values. The invariance
of ranks and singular values can be used for SLOCC\ and LU classifications
of $n$ qubits.

This paper is organized as follows. In the section 2, from the coefficient
matrices of states of n qubits we construct the $\ell $-spin-flipping
matrices and we show that ranks of the $\ell $-spin-flipping matrices are
preserved under SLOCC. In the section 3, we demonstrate how the invariant
ranks are used for SLOCC classification of entanglement. In the section 4,
we argue the invariance of singular values of the $\ell $-spin-flipping
matrices under LU. In the section 5, we investigate the LU classification.

\section{The invariance of ranks of the $\ell $-spin-flipping matrices}

Let $|\psi \rangle =\sum_{i=0}^{2^{n}-1}a_{i}|i\rangle $ be any pure state
of $n$ qubits, where $a_{i}$ are coefficients. It is well known that two $n$%
-qubit pure states $|\psi \rangle $ and $|\psi ^{\prime }\rangle $ are SLOCC
(resp. LU) equivalent if and only if the two states satisfy the following
equation

\begin{equation}
|\psi ^{\prime }\rangle =\mathcal{A}_{1}\otimes \mathcal{A}_{2}\otimes
\cdots \otimes \mathcal{A}_{n}|\psi \rangle ,  \label{SLOCC-1}
\end{equation}%
where $\mathcal{A}_{i}$ are invertible (resp. unitary) \cite{Dur, Kraus2}.

Let $C_{q_{1}\cdots q_{\ell }}(|\psi \rangle )$ be the $2^{\ell }\times
2^{n-\ell }$ coefficient matrix of the state $|\psi \rangle $ of $n$ qubits,
where $q_{1},\cdots ,q_{\ell }$ are the row bits and $q_{\ell +1},\cdots
,q_{n}$ are the column bits. For example, for three qubits,
\begin{equation}
C_{1}(|\psi \rangle )=\left(
\begin{array}{cccc}
a_{0} & a_{1} & a_{2} & a_{3} \\
a_{4} & a_{5} & a_{6} & a_{7}%
\end{array}%
\right) .  \label{M3}
\end{equation}

Let
\begin{eqnarray}
&&\Omega _{q_{1},q_{2},...,q_{i}}^{(n)}(|\psi \rangle )  \notag \\
&=&C_{q_{1},q_{2},...,q_{i}}^{(n)}(|\psi \rangle )\upsilon ^{\otimes
(n-i)}(C_{q_{1},q_{2},...,q_{i}}^{(n)}(|\psi \rangle ))^{T},  \label{mat-1}
\end{eqnarray}%
where $\upsilon =\sqrt{-1}\sigma _{y}$ and $\sigma _{y}\ $is the Pauli
operator.

It is clear that $\Omega _{q_{1},q_{2},...,q_{i}}^{(n)}(|\psi \rangle )$ is
a square matrix of order $2^{i}$ for $n$ qubits. Here, we call $\Omega
_{q_{1},...,q_{i}}^{(n)}(|\psi \rangle )$ a spin-flipping\ matrix. Clearly,
when $n-i$ is even, then $\Omega _{q_{1},...,q_{i}}^{(n)}(|\psi \rangle )$
is symmetric. Otherwise, it is skew-symmetric.

Next, we define the $1$-spin-flipping $\Omega _{q_{1},...,q_{i}}^{(n)}(|\psi
\rangle )^{\odot 1}$ matrix\ as the spin-flipping$\ \Omega
_{q_{1},...,q_{i}}^{(n)}(|\psi \rangle )$\ and the $\ell $-spin-flipping
matrix $\Omega _{q_{1},...,q_{i}}^{(n)}(|\psi \rangle )^{\odot \ell }$\ for $%
\ell >1$ as
\begin{equation}
\Omega _{q_{1},...,q_{i}}^{(n)}(|\psi \rangle )^{\odot \ell }=\Omega
_{q_{1},...,q_{i}}^{(n)}(|\psi \rangle )^{\odot (\ell -1)}\upsilon ^{\otimes
i}\Omega _{q_{1},...,q_{i}}^{(n)}(|\psi \rangle ).  \label{de-2}
\end{equation}

Let $\alpha =\Pi _{k=1}^{i}\det \mathcal{A}_{q_{k}}$ and $\beta =\Pi
_{k=i+1}^{n}\det \mathcal{A}_{q_{k}}$. Then, invoking the fact that $%
\mathcal{A}_{k}^{T}\upsilon \mathcal{A}_{k}=(\det \mathcal{A}_{k})\upsilon $
and from \cite{LDFPRA15}, for two SLOCC equivalent\ states $|\psi ^{\prime
}\rangle $\ and $|\psi \rangle $\ of $n$\ qubits a complicated calculation
yields \

\begin{eqnarray}
&&\Omega _{q_{1},...,q_{i}}^{(n)}(|\psi ^{\prime }\rangle )^{\odot \ell }
\notag \\
=\alpha ^{\ell -1}\beta ^{\ell } &&(\mathcal{A}_{q_{1}}\otimes \cdots
\otimes \mathcal{A}_{q_{i}})\Omega _{q_{1},...,q_{i}}^{(n)}(|\psi \rangle
)^{\odot \ell }\times  \notag \\
&&(\mathcal{A}_{q_{1}}\otimes \cdots \otimes \mathcal{A}_{q_{i}})^{T}.
\label{g-3}
\end{eqnarray}

Eq. (\ref{g-3}) leads to the following theorem.

\textit{Theorem 1. }If two pure states $|\psi ^{\prime }\rangle $\ and $%
|\psi \rangle $\ of $n$\ qubits are SLOCC equivalent, then for any $\ell $
the $\ell $-spin-flipping matrices\ $\Omega _{q_{1},...,q_{i}}^{(n)}(|\psi
^{\prime }\rangle )^{\odot \ell }$\ and $\Omega
_{q_{1},...,q_{i}}^{(n)}(|\psi \rangle )^{\odot \ell }$ are congruent and
then have the same rank.

The contraposition of Theorem 1 says that if the $\ell $-spin-flipping
matrices $\Omega _{q_{1},...,q_{i}}^{(n)}(|\psi ^{\prime }\rangle )^{\odot
\ell }$ and $\Omega _{q_{1},...,q_{i}}^{(n)}(|\psi \rangle )^{\odot \ell }$
have different ranks for some $\ell $, then the two states $|\psi ^{\prime
}\rangle $ and $|\psi \rangle $ belong to different SLOCC classes.

\section{SLOCC classification via ranks of the $\ell $-spin-flipping matrices%
}

Let $r(A)$ be the rank of the matrix $A$. Then, it is well known that $%
r(AB)\leq \min \{r(A),r(B)\}$. Here, $r(\Omega _{q_{1},\cdots
,q_{i}}^{(n)}(|\psi \rangle )^{\odot k})$ is\ denoted as $r_{q_{1},\cdots
,q_{i}}^{(k)}$. Then $r_{q_{1},\cdots ,q_{i}}^{(1)}\geq \cdots \geq
r_{q_{1},\cdots ,q_{i}}^{(k)}\geq \cdots $. If $\Omega _{\
q_{1},...,q_{i}}^{(n)}(|\psi \rangle )$ is not full, then the rank $%
r_{q_{1},\cdots ,q_{i}}^{(k)}$ may decrease as $k$ increases. Therefore, it
is possible that the $\ell $-spin-flipping matrices have different ranks for
some $\ell $ for two SLOCC inequivalent $n$-qubit states. It means that the
theorem 1 may be used for SLOCC\ classification.

For example, for the states GHZ and W of three qubits a simple calculation
shows that $\Omega _{1,2}^{(3)}(|$GHZ$\rangle )$ and $\Omega _{1,2}^{(3)}(|$W%
$\rangle )$ have the same rank 2. But for GHZ, $r_{1,2}^{(2)}=2$ while for
W, $r_{1,2}^{(2)}=1$. So, in light of Theorem 1 GHZ and W are SLOCC
inequivalent.

Next we show how Theorem 1 is used for SLOCC classification.

(1). For any state $|\psi \rangle $ of two qubits, $\Omega _{1}^{(2)}(|\psi
\rangle )=(a_{0}a_{3}-a_{1}a_{2})\upsilon $. It is trivial to see that the
rank of $\Omega _{1}^{(2)}(|\psi \rangle )$ is 2 or 0. Thus, we obtain a
complete classification under SLOCC for two qubits.

(2). By using $r_{1,2}^{(1)}$, $r_{1,2}^{(2)}$, and $r_{1,2}^{(3)}$, a
tedious calculation yields a SLOCC\ classification of three qubits in Table
I.

\begin{table}[tbph]
\caption{The SLOCC classification of three qubits}
\label{table1}%
\begin{ruledtabular}
\begin{tabular}{cccc}
states & $r_{1,2}^{(1)}r_{1,2}^{(2)}r_{1,2}^{(3)}$ & states
& $r_{1,2}^{(1)}r_{1,2}^{(2)}r_{1,2}^{(3)}$ \\ \hline
GHZ & 222 & W & 210 \\
A-BC,B-AC & 200 & C-AB,$|000\rangle $ & 000 \\
\end{tabular}
\end{ruledtabular}
\end{table}

By Table I, we can determine that to which SLOCC\ class a state of three
qubits belongs. For example, let $|\xi \rangle =\frac{1}{2\sqrt{2}}%
(\sum_{i=0}^{6}|i\rangle -|7\rangle )$. Then $r((\Omega _{1,2}^{(3)}(|\xi
\rangle )^{\odot \ell })=2$, $\ell \geq 1$. So, $|\xi \rangle $ belongs to
GHZ class.

(3). By using $r_{1,2}^{(1)}$, $r_{1,2}^{(2)}$, and $r_{1,2}^{(3)}$, we
obtain a complete SLOCC classification of the states of Acin et al.'s
canonical form in Table II.

It is well known that any state of three qubits can be written in the
following Acin et al.'s canonical form:
\begin{eqnarray}
|A\rangle &=&\lambda _{0}|000\rangle +\lambda _{1}e^{i\varphi }|100\rangle
\notag \\
&&+\lambda _{2}|101\rangle +\lambda _{3}|110\rangle +\lambda _{4}|111\rangle
,  \label{Acin-1-}
\end{eqnarray}%
where $\lambda _{i}\geq 0$, $i=0,1,2,4$, $0\leq \varphi \leq \pi $, and $%
\sum_{i=0}^{4}\lambda _{i}^{2}=1$ \cite{Acin00, Acin01}. Via $r_{1,2}^{(1)}$%
, $r_{1,2}^{(2)}$, $r_{1,2}^{(3)}$, and Table I, a tedious and
straightforward calculation yields the complete SLOCC\ classification of the
states of the Acin's canonical form in Table II. Thus, we recover D\"{u}r et
al.'s six SLOCC\ classes obtained via local ranks \cite{Dur}. Here, each one
of the six classes is parametrized by parameters $\lambda _{1}$, $\lambda
_{2}$, $\lambda _{3}$, and $\lambda _{4}$. For example, the GHZ class is
described by $\lambda _{0}\lambda _{4}\neq 0$.\ For another example, we name
as the SLOCC\ W class the SLOCC class in Table II characterized by the
parameters $\lambda _{0}\neq 0$, $\lambda _{4}=0$, and $\lambda _{2}\lambda
_{3}\neq 0$. We can show that the W state belongs to the SLOCC W class
below. Let $|\vartheta \rangle =\frac{1}{\sqrt{3}}(|000\rangle +|101\rangle
+|110\rangle )$. One can test that $\sigma _{x}\otimes I\otimes I|\vartheta
\rangle =|$W$\rangle $ and $|\vartheta \rangle $\ belongs to the SLOCC W
class. \

\begin{table}[tbph]
\caption{Complete SLOCC classification of states of Acin et al.'s canonical
form}
\label{table2}%
\begin{ruledtabular}

\begin{tabular}{cccc}
&  & $r_{1,2}^{(1)}r_{1,2}^{(2)}r_{1,2}^{(3)}$ & classes \\ \hline
$\lambda _{0}=0,\lambda _{4}\neq 0$
& $\lambda _{2}\lambda _{3}=\lambda_{1}\lambda _{4}e^{i\varphi }$ & $000$ & A-B-C \\
& $\lambda _{2}\lambda _{3}\neq \lambda _{1}\lambda _{4}e^{i\varphi }$ & $200
$ & A-BC \\
$\lambda _{0}\neq 0,\lambda _{4}=0$ & $\lambda _{2}=\lambda _{3}=0$ & $000$
& A-B-C \\
& $\lambda _{2}=0,\lambda _{3}\neq 0$ & $000$ & C-AB \\
& $\lambda _{2}\neq 0,\lambda _{3}=0$ & $200$ & B-AC \\
& $\lambda _{2}\lambda _{3}\neq 0$ & $210$ & W \\
$\lambda _{0}=\lambda _{4}=0$ & $\lambda _{2}\lambda _{3}=0$ & $000$ & A-B-C
\\
& $\lambda _{2}\lambda _{3}\neq 0$ & $200$ & A-BC \\
$\lambda _{0}\lambda _{4}\neq 0$ &  & $222$ & GHZ \\

\end{tabular}
\end{ruledtabular}
\end{table}

\section{ The invariance of singular values of the $\ell $-spin-flipping
matrices\ under LU}

Here, two matrices $A$ and $B$ are called unitary congruent if there is a
unitary matrix $P$ such that $B=PAP^{T}$. When two states are LU equivalent,
$\mathcal{A}_{i}$ in Eq. (\ref{SLOCC-1})\ are unitary. Thus, Eq. (\ref{g-3})
leads to the following theorem.

\textit{Theorem 2}. If two pure states $|\psi ^{\prime }\rangle $ and $|\psi
\rangle $\ of $n$\ qubits are LU equivalent, then the $\ell $-spin-flipping
matrices$\ \Omega _{\ q_{1},...,q_{i}}^{(n)}(|\psi ^{\prime }\rangle
)^{\odot \ell }$\ and $\Omega _{\ q_{1},...,q_{i}}^{(n)}(|\psi \rangle
)^{\odot \ell }$ are unitary congruent, and then

(1) have the same ranks,

(2) have the same singular values, and

(3) $\Omega _{\ q_{1},...,q_{i}}^{(n)}(|\psi ^{\prime }\rangle )$ and $%
\Omega _{\ q_{1},...,q_{i}}^{(n)}(|\psi \rangle )$ have the same absolute
values of determinants.

First we demonstrate how to partition Verstraete et al.'s nine families
under LU invoking absolute values of determinants. For example, for the
family $L_{a_{4}}$ \cite{Verstraete}, $|\det \Omega _{\
1,2}^{(4)}(|L_{a_{4}}\rangle )|=|a^{4}|^{2}$. Thus, two states of the family
$L_{a_{4}}$ with different values of $|a|$ are different under LU.
Similarly, under LU\ we can partition the families $G_{abcd}$, $L_{abc_{2}}$%
, $L_{a_{2}b_{2}}$, and $L_{ab_{3}}$ in \cite{Verstraete}.

Next we use the following example to explain that the invariance of singular
values is more powerful than the invariance of the ranks for LU
classification. Let us consider two states $|W_{1}\rangle $ and $%
|W_{2}\rangle $ of the SLOCC W class of three qubits in Table II, where $%
|W_{1}\rangle =\frac{1}{2}(|000\rangle +|100\rangle +|101\rangle +|110))$
and $|W_{2}\rangle =\lambda _{0}|000\rangle +\lambda _{2}|101\rangle
+\lambda _{3}|110\rangle $, where $\lambda _{0}^{2}=\frac{1}{4}(\frac{1}{2}-%
\frac{\sqrt{2}}{4})$, $\lambda _{2}^{2}=\frac{1}{2}+$ $\frac{\sqrt{2}}{4}$,
and $\lambda _{3}^{2}=\frac{3}{4}(\frac{1}{2}-\frac{\sqrt{2}}{4})$. A
calculation makes Table III. From Table III, clearly we cannot distinguish $%
|W_{1}\rangle $ and $|W_{2}\rangle $ under LU via the ranks $r_{1,2}^{(1)}$,
$r_{1,2}^{(2)}$, and $r_{1,2}^{(3)}$ or the singular values $\sigma _{1}$, $%
\sigma _{2}$, $\sigma _{3}$, and $\sigma _{4}$ of $\Omega
_{1,2}^{(3)}(|W_{i}\rangle )$. Whereas, we can distinguish them under LU\
via the singular values $\eta _{1}$, $\eta _{2}$, $\eta _{3}$, and $\eta
_{4} $\ of $\Omega _{1,2}^{(3)}(|W_{i}\rangle )^{\odot 2}$.

\begin{table}[tbph]
\caption{Two LU inequivalent states of three qubits}
\label{table3}%
\begin{ruledtabular}
\begin{tabular}{cccc}

state & $r_{1,2}^{(1)}r_{1,2}^{(2)}r_{1,2}^{(3)}$ & $\sigma _{1}^{2}\sigma
_{2}^{2}\sigma _{3}^{2}\sigma _{4}^{2}$ & $\eta _{1}^{2}\eta _{2}^{2}\eta
_{3}^{2}\eta _{4}^{2}$ \\ \hline
$|W_{1}\rangle $ & 210 & $\frac{1}{8}\frac{1}{8}00$ & $\frac{1}{4096}\frac{1%
}{4096}00$ \\
$|W_{2}\rangle $ & 210 & $\frac{1}{8}\frac{1}{8}00$ & $\frac{3}{16384}\frac{3%
}{16384}00$ \\

\end{tabular}
\end{ruledtabular}
\end{table}

\section{LU classification of $n$ qubits via singular values of\ the $\ell $%
-spin-flipping matrices}

\subsection{LU classification of even $n$ qubits invoking singular values of
$\Omega _{1}^{(n)}$}

\subsubsection{Singular values of $\Omega _{1}^{(n)}$ are just the
concurrence for even $n$ qubits}

For any state $|\psi \rangle $ of even $n$ qubits, let $t_{1}$ and $t_{2}$
be the singular values of the skew-symmetric matrix$\ \Omega
_{1}^{(n)}(|\psi \rangle )$. Then, a calculation yields that
\begin{equation}
t_{1}=t_{2}=\left\vert
\sum_{i=0}^{2^{n-1}-1}(-1)^{N(i)}a_{i}a_{2^{n}-i-1}\right\vert .
\label{con-2}
\end{equation}%
Here $N(i)$ is the number of 1s in the $n$-bit binary representation $%
i_{n-1}...i_{1}i_{0}$ of $i$. That is, $N(i)$ is the parity of $i$. It is
known that the singular values of $\Omega _{1}^{(n)}(|\psi \rangle )$ in Eq.
(\ref{con-2}) are just the concurrence of even $n$ qubits for the state $%
|\psi \rangle $ \cite{LDFPRA13}. For two qubits, $t_{1}=t_{2}=\left\vert
a_{0}a_{3}-a_{1}a_{2}\right\vert $.

\subsubsection{LU classification of even $n$ qubits invoking the invariance
of the concurrence for even $n$ qubits}

Theorem 2 and Eq. (\ref{con-2}) lead to the following theorem.

\textit{Theorem 3.} If two pure states of even $n$ qubits are LU equivalent
then the two states have the same concurrence for even $n$ qubits.

In comparison, if two pure states of even $n$ qubits are SLOCC equivalent
then either their concurrences for even $n$ qubits both vanish or neither
vanishes \cite{LDFPRA13}.

From Theorem 3, if two states of even $n$ qubits have different
concurrences, then they belong to different LU equivalence classes. For
example, for two qubits the Bell states $\frac{1}{\sqrt{2}}(|00\rangle
+|11\rangle )$ have the maximal concurrence $\frac{1}{2}$. let $|\zeta
\rangle =$ $\frac{1}{\sqrt{3}}|00\rangle +\frac{\sqrt{2}}{\sqrt{3}}%
|11\rangle $. The concurrence of $|\zeta \rangle $ is $\frac{\sqrt{2}}{3}$.
In light of Theorem 3, $|\zeta \rangle $ is LU inequivalent to the Bell
state though $|\zeta \rangle $ is SLOCC equivalent to the Bell state. For
two qubits, the Schmidt coefficients are used for LU classification \cite%
{Kraus1}.

Let $c$ denote the concurrence. Clearly, $0\leq c\leq \frac{1}{2}$. And let
the family, denoted as $F_{c}$, consist of all\ the states with the same
concurrence $c$. It means that if two states are LU equivalent then they
belong to the same family $F_{c}$. For example, GHZ and $|0\cdots 0\rangle $
of any even $n$ qubits belong to $F_{1/2}$ and $F_{0}$, respectively.\
Clearly, for any even $n$ qubits there is a one to one correspondence
between the set $\{F_{c}|c\in \lbrack 0,1/2]\}$\ of the families $F_{c}$\
and the interval $[0,1/2]$.

\subsection{LU classification of odd $n$ qubits invoking singular values of $%
\Omega _{1}^{(n)}$}

\subsubsection{A product of singular values of $\Omega _{1}^{(n)}(|\protect%
\psi \rangle )$ is just the n-tangle for odd $n$ qubits}

For any state $|\psi \rangle $ of odd $n$ qubits, the symmetric matrix $%
\Omega _{1}^{(n)}(|\psi \rangle )$ can be written as $\left(
\begin{array}{cc}
e_{11} & e_{12} \\
e_{12} & e_{22}%
\end{array}%
\right) $. We calculate $e_{11}$, $e_{12}$, and $e_{22}$ in Appendix A. Let $%
t_{1}$ and $t_{2}$ be the singular values of $\Omega _{1}^{(n)}(|\psi
\rangle )$. Then, a calculation yields that $t_{1}^{2}=\frac{\Delta +\sqrt{%
\Delta ^{2}-4D}}{2}$ and $t_{2}^{2}=\frac{\Delta -\sqrt{\Delta ^{2}-4D}}{2}$%
, where $\Delta =2|e_{12}|^{2}+|e_{11}|^{2}+|e_{22}|^{2}$ and $%
D=|e_{11}e_{22}-e_{12}^{2}|^{2}$.

\subsubsection{LU classification of odd $n$ qubits invoking the invariance
of the n-tangle for odd $n$ qubits}

Note that $t_{1}^{2}+t_{2}^{2}=\Delta $ and $%
t_{1}t_{2}=|e_{11}e_{22}-e_{12}^{2}|$. Recall that $%
|e_{11}e_{22}-e_{12}^{2}| $ is just the n-tangle of odd $n$ qubits \cite%
{LDFQIP, LDFPRA15}. Thus, the invariance of singular values of $\Omega
_{1}^{(n)}(|\psi \rangle )$ implies that the n-tangle and $\Delta $ are
invariant under LU. For example, the above states $|W_{1}\rangle $ and $%
|W_{2}\rangle $ are different in $\Delta $, so they are LU inequivalent.

We can conclude the following theorem from Theorem 2 and the above
discussion.

\textit{Theorem 4.} If two pure states of odd $n$ qubits are LU equivalent
then they have the same n-tangle for odd $n$ qubits.

In comparison, if two pure states of odd $n$ qubits are SLOCC equivalent
then either their n-tangles for odd $n$ qubits both vanish or neither
vanishes \cite{LDFQIP}.

From Theorem 4, one can see that two pure states of odd $n$ qubits are LU
inequivalent provided that the two states are different in n-tangle. For $|$%
GHZ$\rangle $, the n-tangle is $1/4$, so any state whose n-tangle is not $%
1/4 $ is LU inequivalent to $|$GHZ$\rangle $. It is known that the n-tangle
of odd $n$ qubits is between 0 and $1/4$. We can define the family $F_{g}$
to be the set of all the states whose n-tangles are $g$. Thus, if two pure
states of odd $n$ qubits are LU equivalent then they belong to the same
family $F_{g}$. It is plain to see that for any odd $n$ qubits, there is a
one to one correspondence between the set $\{F_{g}|g\in \lbrack 0,1/4]\}$ of
the families $F_{g}$ and the interval $[0,1/4]$.

\subsection{LU classification of three qubits invoking singular values of $%
\Omega _{1,2}^{(3)}$}

\subsubsection{Singular values of $\Omega _{1,2}^{(3)}$}

A straightforward calculation yields that the singular values of $\Omega
_{1,2}^{(3)}(|\psi \rangle )$\ are $S$, $S$, $0$, $0$, where $S$ is put in
Appendix A. For any state $|A\rangle $\ of Acin et al.'s canonical form in
Eq. (\ref{Acin-1-}), $S^{2}$ reduces to

\begin{equation}
S^{2}=\lambda _{0}^{2}\lambda _{2}^{2}+\lambda _{0}^{2}\lambda
_{4}^{2}+|\lambda _{1}\lambda _{4}e^{i\varphi }-\lambda _{2}\lambda
_{3}|^{2}.  \label{Acin-3}
\end{equation}

Specially, for the SLOCC GHZ class in Table II, we have $S$ in Eq. (\ref%
{Acin-3}); for the SLOCC W class in Table II, $S^{2}=\lambda
_{2}^{2}(\lambda _{0}^{2}+\lambda _{3}^{2})$; for the SLOCC A-BC class, $%
S=|\lambda _{1}\lambda _{4}e^{i\varphi }-\lambda _{2}\lambda _{3}|$; for the
SLOCC B-AC\ class, $S=\lambda _{0}\lambda _{2}$; for the SLOCC C-AB class
and the SLOCC $|000\rangle $ class, $S=0$.

\subsubsection{LU classification of Acin et al.'s\ canonical form for three
qubits invoking singular values of $\Omega _{1,2}^{(3)}$}

Recall that for three qubits, the space of the normalized states in Eq. (\ref%
{Acin-1-}) is partitioned into nine families under LU \cite{Acin00, Acin01}.
It is well known that pure states of three qubits were partitioned into six
SLOCC\ classes: GHZ, W, A-BC, B-AC, C-AB, and A-B-C \cite{Dur}. In terms of
the singular values, we partition each one of the SLOCC\ classes GHZ, W,
A-BC, B-AC, and C-AB in Table IV\ under LU as follows.

In light of Theorem 4,\ one can know that two states are LU inequivalent if
the two states are different in $S$. Next, let us demonstrate how to
partition the SLOCC\ class W under LU. Let the family $F_{S}$ consist of
all\ the states of the SLOCC W\ class with the same value of $S$. Thus, we
obtain a one to one correspondence between the set $\{F_{s}|S\in (0,\sqrt{2}%
/3]\}$\ of the LU families $F_{s}$ and the interval $(0,\sqrt{2}/3]$.
Similarly, we can partition GHZ, B-AC, and A-BC under LU. See Table IV.

For the SLOCC C-AB class, $S=0$. Note that Acin et al.'s canonical form for
SLOCC C-AB class in Table II\ reduces to $|\varsigma \rangle _{AB}|0\rangle
_{C}$, where $|\varsigma \rangle _{AB}=(\lambda _{0}|00\rangle +\lambda
_{1}e^{i\varphi }|10\rangle +\lambda _{3}|11\rangle )_{AB}$, where $\lambda
_{0}\lambda _{3}\neq 0$. Let $c$ be the concurrence of $|\varsigma \rangle
_{AB}$ and the family $F_{c}$ consist of all the states $|\varsigma \rangle
_{AB}|0\rangle _{C}$ with the same value in $c=\lambda _{0}\lambda _{3}$.
Thus, there is a one to one correspondence between the set $\{F_{c}|c\in
(0,1/2]\}$ of the families $F_{c}$ and the interval $(0,1/2]$.

\begin{table}[tbph]
\caption{LU classification of the states of Acin et al.'s canonical form}
\label{table4}%
\begin{ruledtabular}

\begin{tabular}{cc}

SLOCC\  & the set of LU families \\ \hline
GHZ class & $\{F_{s}|S\in (0,1/2]\}$ \\
W class & $\{F_{s}|S\in (0,\sqrt{2}/3]\}$ \\
B-AC class & $\{F_{s}|S\in (0,1/2]\}$ \\
A-BC class & $\{F_{s}|S\in (0,1/2]\}$ \\
C-AB class & $\{F_{c}|c\in (0,1/2]\}$ \\
$|000\rangle $ class & single family \\

\end{tabular}
\end{ruledtabular}
\end{table}

\section{Summary}

In this paper, from the coefficient matrices of states of n qubits we
construct the $\ell $-spin-flipping matrices and show that the $\ell $%
-spin-flipping matrices are congruent and unitary congruent under SLOCC and
LU, respectively. Thus, the ranks and the singular values of the $\ell $%
-spin-flipping matrices are invariant under SLOCC and LU, respectively.

The invariance of ranks of the spin-flipping matrices\ provides a simple way
of classifying $n$-qubit states under SLOCC. For example, we obtain complete
SLOCC\ classifications of two and three qubits. The invariance of singular
values of the spin-flipping matrices$\ \Omega _{1}^{(n)}$\ implies the
invariance of the concurrence for even $n$ qubits and the invariance of the
n-tangle for odd $n$ qubits. The invariance of the concurrence and the
invariance of the n-tangle permit LU\ classifications for even $n$ qubits
and odd $n$ qubits, respectively. See Table V. It only performs additions
and multiplications of coefficients of states to compute the concurrence for
even $n$ qubits and the n-tangle for odd $n$ qubits in comparison to the
methods \cite{Kraus1, Kraus2}.

\begin{table}[tbph]
\caption{SLOCC and LU classification invoking the concurrence and n-tangle}
\label{table5}%
\begin{ruledtabular}
\begin{tabular}{cc}

qubits & $\psi ^{\prime }$ and $\psi $\ are SLOCC inequivalent \\ \hline
even $n$ & if only one of their concurrences is 0 \\
odd $n$ & if only one of their n-tangles is 0 \\ \hline \hline
qubits & $\psi ^{\prime }$ and $\psi $\ are LU inequivalent \\ \hline
even $n$ & if their concurrences are different \\
odd $n$ & if their n-tangles are different \\

\end{tabular}
\end{ruledtabular}
\end{table}

Acknowledgement---This work was supported by NSFC (Grant No. 10875061) and
Tsinghua National Laboratory for Information Science and Technology.

\section{Appendix A. Some expressions}

\begin{equation*}
e_{12}=\sum_{i=0}^{2^{n-1}-1}(-1)^{N(i)}a_{i}a_{2^{n}-1-i},
\end{equation*}

\begin{equation*}
e_{11}=2\sum_{i=0}^{2^{n-2}-1}(-1)^{N(i)}a_{i}a_{2^{n-1}-1-i},
\end{equation*}%
\begin{equation*}
e_{22}=2\sum_{i=0}^{2^{n-2}-1}(-1)^{N(i)}a_{2^{n-1}+i}a_{2^{n}-1-i}.
\end{equation*}

\begin{eqnarray*}
S^{2} &=&|a_{0}a_{3}-a_{1}a_{2}|^{2}+|a_{0}a_{5}-a_{1}a_{4}|^{2} \\
&&+|a_{0}a_{7}-a_{1}a_{6}|^{2}+|a_{2}a_{5}-a_{3}a_{4}|^{2} \\
&&+|a_{2}a_{7}-a_{3}a_{6}|^{2}+|a_{4}a_{7}-a_{5}a_{6}|^{2}.
\end{eqnarray*}

\end{document}